\documentclass[12pt]{iopart}

\usepackage{graphicx}
\usepackage{color}

\begin{document}

\title[Experimental quantum tossing of a single coin]{Experimental quantum tossing of a single coin}

\author{A. T. Nguyen$^1$, J. Frison$^2$, K. Phan Huy$^3$ and S. Massar$^2$}

\address{$^1$ Service OPERA photonique, CP 194/5, Universit\'{e} Libre
de Bruxelles (U.L.B.), Avenue F. D. Roosevelt 50, B-1050
Bruxelles, Belgium }

\address{$^2$ Laboratoire d'Information Quantique, CP 225,
 Universit\'{e} Libre de Bruxelles (U.L.B.),
 Boulevard du Triomphe, B-1050 Bruxelles, Belgium}

\address{$^3$ D\'{e}partement d'Optique P.M. Duffieux, Institut
FEMTO-ST, Centre National de la Recherche Scientifique UMR 6174,
Universit\'{e} de Franche-Comt\'{e}, 25030 Besan{\c c}on, France}

\eads{\mailto{annguyen@ulb.ac.be}, \mailto{smassar@ulb.ac.be}}

\begin{abstract}
The cryptographic protocol of coin tossing consists of two
parties, Alice and Bob, that do not trust each other, but want to
generate a random bit. If the parties use a classical
communication channel and have unlimited computational resources,
one of them can always cheat perfectly. Here we analyze in detail
how the performance of a quantum coin tossing experiment should be
compared to classical protocols, taking into account the
inevitable experimental imperfections. We then report an
all-optical fiber experiment in which a single coin is tossed
whose randomness is higher than achievable by any classical
protocol and present some easily realisable cheating strategies by
Alice and Bob.

\end{abstract}

\pacs{42.50.-p,42.50.Ex}

%\submitto{NJP}

\maketitle

\section{Introduction}
The cryptographic protocol of coin tossing introduced by Blum
\cite{Blum} consists of two parties, Alice and Bob, that do not
trust each other, but want to generate a random bit. If the
parties use a classical communication channel and have unlimited
computational resources, one of them can always cheat perfectly.
But what if they use a quantum communication channel? Because of
its conceptual importance and potential applications, quantum coin
tossing was already envisaged by Bennett and Brassard in their
seminal paper on quantum cryptography \cite{BB84}. Later works
showed that perfect quantum coin tossing is impossible
\cite{LC,MSCK,Kitaev}, but that imperfect protocols exist
\cite{GVW,MSCK,ATSV,A2002,SR2002,Mo07} that perform better than
any classical protocol.

Work on quantum coin tossing distinguishes between ``weak coin
tossing" and ``strong coin tossing".  In weak coin tossing Alice
and Bob have antagonistic goals: Alice wants the coin to be heads,
say, whereas Bob wants the coin to come out tails. Good quantum
protocols for weak coin tossing exist, although they seem very
difficult to implement\cite{Mo07}. In strong coin tossing Alice
and Bob both want the coin to be perfectly random. Quantum
protocols that perform better at strong coin tossing than any
classical protocol exist\cite{A2002,SR2002} and come close to the
known upper bound (for the original unpublished proof of the
upperbound, see\cite{Kitaev}; published proofs can be found in
\cite{ABDR,GW}).

Quantum coin tossing itself is just one example of several
interesting tasks that two parties which do not trust each other
can achieve if they share a quantum communication channel, but
cannot achieve if they use a classical communication channel.
Other examples include multiparty coin tossing\cite{ABDR} and weak
forms of string committment\cite{BCHLW,J}. The no go theorems
mentioned above \cite{LC,MSCK,Kitaev} rule out most other
applications, except if one adds additional assumptions such as
bounding the size of quantum memories\cite{DFSS}.

 Recently two works
\cite{Expt2,Expt} have experimentally studied optical
implementations of quantum coin tossing. However the experiment of
ref. \cite{Expt2} suffered from important photon loss which made
it difficult to assess how the experiment worked when tossing a
single coin.  This was circumvented, as in \cite{Expt}, by
addressing  string flipping, {\it i.e.} the problem where the
parties try to toss a string of coins rather than a single one.
These works were however carried out without realizing that good
classical protocols exist for string flipping, see e.g.
\cite{Betal} for a presentation of such protocols.

In the present work we go back to the conceptually simpler problem
of tossing a single coin, and report an experiment in which a
single coin is tossed whose randomness is higher than achievable
by any classical protocol. We begin by discussing in detail how
the results of such a coin tossing experiment should be compared
with classical protocols in view of the inevitable imperfections
that will occur in any experimental realisation. Coin tossing in
the presence of noise was already studied in \cite{BM}, but with
the emphasiz on applications to string flipping, whereas here we
are concerned with tossing a single coin. We then present the
experimental implementation, which follows closely the earlier
work of \cite{Expt}, and present some easily realisable cheating
strategies by Alice and Bob.

\section{Formulation of the problem}
A protocol for coin tossing consists in a series of rounds of
(classical or quantum) communication at the end of which the
parties decide on an outcome. The outcome can be either a decision
that the coin has the value $c=0$ or $c=1$, or it can be that the
protocol aborts, in which case we say that $c=\perp$.
%In an ideal coin tossing protocol, the outcomes  $c=0$ and $c=1$ both occur with
% probability $1/2$ when the parties are honest. And if one party tries to cheat, then with
%high probability  $c=\perp$, ie. the cheater is caught.
%However because of experimental imperfections, we expect the
%outcome $c=\perp$ to occur even  when both parties are honest. And
%because of the limitations inherent to quantum mechanics and of
%experimental imperfections we expect the probability that a
%cheater to successfully influence the outcome without being caught
%to be quite high.
Note that because the rounds of (quantum or classical)
communication are sequential, it is logically possible for Alice
to choose one output $x$, and for Bob to chose another output $y$.
For the sake of generality it is convenient to take this into
account and to denote by
\begin{eqnarray}
p_{xy}&=&\mbox{Probability that in an honest execution of the}\nonumber\\
& &\mbox{protocol
  Alice outputs $x$ and Bob outputs $y$,} \nonumber\\
& &\mbox{where $x,y\in\{0,1,\perp\}$.}\nonumber
\end{eqnarray}

We will say that a protocol is {\em correct}, if, when both
parties are honest, at the end of the protocol they agree on the
outcome, and that the results $c=0$ and $c=1$ occur with equal
probability: $p_{00}=p_{11}=(1-p_{\perp \perp})/2$. This
formulation takes into account that because of experimental
imperfections, the outcome $c=\perp$ may occur even when both
parties are honest.

The aim of a cheater is to force the outcome of the coin tossing
protocol. We denote by
\begin{eqnarray}
p_{*y}&=&\mbox{Probability that a dishonest Alice can force
 an}\nonumber\\
&& \mbox{ honest Bob to
  output y}\nonumber\\
p_{x*}&=&\mbox{Probability that a dishonest Bob can force an}\nonumber\\
&& \mbox{honest Alice to
  output x}\nonumber
\end{eqnarray}
An alternative notation often used in the litterature is the bias
$\epsilon$ which is related to our notation by
\begin{eqnarray}
\epsilon_A &=& \max_y (p_{*y}-\frac{1}{2} )\nonumber\\
\epsilon_B &=& \max_x (p_{x*}-\frac{1}{2} )
\end{eqnarray}
The bound due to Kitaev \cite{Kitaev,ABDR,GW} states that either
$\epsilon_A$ or $\epsilon_B$ is greater or equal to $1/\sqrt{2}$.
The best known protocol for strong coin tossing due to Ambainis
has $\epsilon_A = \epsilon_B = 1/4$. In our experimental
implementation, as we will see later, we will be concerned by a
protocol which in the terminology of \cite{SR2002} has ``$\rho_0$
and $\rho_1$ both pure''. For such protocols it is proven in
\cite{SR2002} that $\epsilon_A^2 + \epsilon_B^2 \geq 1/4$.

In the appendix we prove the following (which generalises a result
of \cite{Kitaev} when $p_{\perp\perp}=0$):
\\
{\em Lemma 1: For any correct {\em classical} coin tossing
protocol with three outcomes $0,1,\perp$ we have:
\begin{eqnarray}
(1-p_{0*})(1-p_{*1})&\leq& p_{\perp\perp} \ ,\label{r3}\\
(1-p_{1*})(1-p_{*0})&\leq& p_{\perp\perp}\ . \label{r4}
\end{eqnarray}
%where $p_{\perp\perp}$ is the probability that the protocol aborts when both
%parties are honest.
}

Note that if $p_{\perp\perp}=0$ these inequalities imply that
either $p_{0*}=1$ or $p_{*1}=1$, and that either $p_{1*}=1$ or
$p_{*0}=1$, thereby showing that classical coin tossing is
impossible. When $p_{\perp\perp}\neq 0$ a cheater can no longer
necessarily force  the outcome he wants. In the supplementary
material we show that there exist classical protocols that
saturate either one of equations (\ref{r3}) or (\ref{r4}), and
that there exist classical protocols that come close to saturating
both equations (\ref{r3}) and (\ref{r4}).

In view of Lemma 1, it is natural to quantify the quality of
quantum coin tossing experiments by the following merit function:
\begin{equation}
{\cal M} = \frac{ (1-p_{*0})(1-p_{1*})}{2} +
\frac{(1-p_{*1})(1-p_{0*})}{2} - p_{\perp \perp} \label{M}
\end{equation}
which has the following properties:
\begin{enumerate}
\item Positivity of probabilities implies  $-1 \leq {\cal M} \leq
+1$ \item For any classical protocol we have ${\cal M}\leq 0$
\item An ideal protocol would have ${\cal M}=1$.
\end{enumerate}
The interpretation of the merit function is most obvious in the
weak coin tossing scheme wherein Alice  wins if Bob outputs $1$
while Bob wins if Alice outputs $0$ because then the term
$(1-p_{*1})(1-p_{0*})$ is the product of how often a dishonest
Alice cannot force a win times how often a dishonest Bob cannot
force a win (and similarly for the term $(1-p_{*0})(1-p_{1*})$).
The better the protocol, the larger these terms.

As illustration let us compute the value of ${\cal M}$ for
different protocols. The bound due to Kitaev states with
precision, see \cite{ABDR}, that $p_{*1}p_{1*}\geq 1/2$ and
$p_{*0}p_{0*}\geq 1/2$. Inserting this into eq . (\ref{M}) shows
that for all quantum protocols, ${\cal M}\leq (1 -
1/\sqrt{2})^2\simeq 0.086$. For Ambainis's protocol \cite{A2002}
for instance we have ${\cal M}= 1/16=0.0625$.

\section{Experimental Implementation}

\subsection{The Protocol}

Our implementation of quantum coin tossing uses the following
protocol:
\begin{enumerate}
\item Alice chooses $a\in\{0,1\}$ at random. She prepares state
$\psi_a$, where the two possible  states are non orthogonal: $ |
\langle \psi_1 | \psi_0\rangle | = \cos \theta > 0$. She sends
$\psi_a$ to Bob.

The states $\psi_{0,1}$ will be taken to be coherent states of
light of amplitude $\alpha$ and opposite phase:
\begin{equation}
|\psi_{0}\rangle = |+\alpha\rangle \quad , \quad |\psi_{1}\rangle
= |-\alpha\rangle
\end{equation}
which implies that
\begin{equation}
\cos^2 \theta = |\langle\psi_1|\psi_0\rangle|^2= |\langle -\alpha
| + \alpha \rangle |^2 = e^{- 4 \alpha^2}.
\label{cos}\end{equation} In the notation of \cite{SR2002} we thus
have $\rho_0=|\psi_{0}\rangle\langle\psi_{0}|$ and
$\rho_1=|\psi_{1}\rangle\langle\psi_{1}|$ both pure. Also note
that $\rho_0\neq\rho_1$ prevents from cheating strategies based on
entanglement \cite{M96,LC}.

\item Bob chooses $b\in\{0,1\}$ at random. He tells the value of
$b$ to Alice.

\item Alice tells Bob the value of $a$.

\item Bob carries out a measurement which projects onto $\psi_a$
or onto the orthogonal space.
 If he finds that the state is not  equal to $\psi_a$ he aborts,
 and the outcome of  the protocol  is  $\perp$. If he finds that the
 state is equal to $\psi_a$ then the outcome of the protocol is $c=a\oplus b$.

Bob's measurement is carried out as follows: using a local
oscillator (LO), he displaces the quantum state by $+\alpha$ if
$a=1$ or by $-\alpha$ if $a=0$.
 If Alice is honest this results in the state becoming the vacuum state.
 To check this Bob then sends the resulting state onto a single photon detector.
 If the detector clicks then Bob assumes that Alice was cheating and he aborts:
 the outcome of the protocol is $\perp$. If the detector does not click, then Bob assumes
 that Alice is honest. (Note that Bob's measurement
 is similar in spirit to the method proposed in \cite{WV} for quantum state tomography, but Bob's task
 is simpler since he only needs to detect if Alice is cheating, and not carry out the full state tomography).

\end{enumerate}

\subsection{Analysis in the absence of imperfections}
We now study how the merit function ${\cal M}$ depends on the
details of the experiment. For the sake of comparison we first
look at the situation in the absence of imperfections.

First of all, in this case $p_{\perp \perp}=0$.

Second, if Alice is dishonest she will send a fixed state
$|\phi\rangle$ at step 1  and at step 3 she will choose the value
of $a$ which will make her win the protocol, and then she will
hope that Bob will not abort. The probability that Bob will abort
is given by the overlap of $|\phi\rangle$ with $| \psi_0\rangle$
and $| \psi_1\rangle$. One easily finds (see \cite{BM}) that
Alice's optimal choice is $|\phi\rangle = N (| \psi_0\rangle + |
\psi_1\rangle ) $ where $N$ a normalization constant, yielding the
optimal values:
\begin{equation}
p_{*0} =  p_{*1} = \frac{1}{2} + \frac{ |\langle
\psi_1|\psi_0\rangle| }{2} = \frac{1}{2} + \frac{\cos \theta}{2}\
. \label{p*c}
\end{equation}

Third, if Bob is dishonest, he will measure the state sent by
Alice at step 2  so as to try to find out whether it is $\psi_0$
or $\psi_1$, and he will then choose the value of $b$ according to
the result of his measurement. For the optimal measurement the
probability that Bob wins is
\begin{equation}
p_{0*} =  p_{1*} = \frac{1}{2} + \frac{ \sqrt{1- |\langle
\psi_1|\psi_0\rangle|^2} }{2} = \frac{1}{2} + \frac{\sin
\theta}{2}\ . \label{pc*}
\end{equation}

The maximal value of the merit function ${\cal M}_{max}= \frac{ (1
- 1/\sqrt{2})^2}{4}\simeq 0.021$ occurs when
$\cos(\theta)=\sin(\theta)=1/\sqrt{2}$, corresponding to
$\alpha^2=0.17$. Note that this is the maximum value for protocols
which in the terminology of Spekkens and Rudolph \cite{SR2002}
fall in the category ``$\rho_0$ and $\rho_1$ both pure''.

\subsection{Analysis in the presence of imperfections}
To obtain estimates on $p_{*c}$, $p_{c*}$ and $p_{\perp\perp}$,
and hence to estimate ${\cal M}$, in the presence of imperfections
requires that we make assumptions on how the experiment is carried
out.

%\subsubsection{Both parties are honest}

The parameter, $p_{\perp\perp}$, which we also call the Quantum
Bit Error Rate (QBER), can easily be measured experimentally by
tossing a large number of coins with Alice and Bob both following
their honest strategy.

\subsubsection{Bob is dishonest}
When Bob is dishonest his cheating strategy is, as before, to
estimate before step 2   the state $|\psi_{a} \rangle$ prepared by
Alice so as to correctly guess the value of $a$. How do
experimental imperfections, and in particular the limited
visibility $V$ of interferences affect Bob's success probability
$p_{c*}$? To analyse this note that the state Alice sends to Bob
is a short laser pulse of known intensity which is then strongly
attenuated. Under strong attenuation all quantum states tend
towards mixtures of coherent states (see e.g. \cite{Expt}). Thus
we can assume that the states prepared by Alice are coherent
states of known intensity $\alpha^2$. These coherent states are
not precisely known to Alice. However it is not difficult to show
that if two coherent states have intensity $\alpha^2$, their
scalar product is lower bounded by $|\langle \psi_1|\psi_0\rangle|
\geq e^{-2 \alpha^2}$. Bob's cheating probability can then be
bounded, as in equation (\ref{pc*}), by the scalar product of the
two states prepared by Alice:
\begin{equation}
p_{c*}=\frac{1}{2} +
 \frac{ \sqrt{1- |\langle \psi_1|\psi_0\rangle|^2} }{2} \leq \frac{1}{2} +
                \frac{
            \sqrt{1- e^{-4 \alpha^2} }
                }{2}\ .
\label{pc*2}
\end{equation}

\subsubsection{Alice is dishonest}
When Alice is dishonest we  suppose that she can prepare an
arbitrary state   just in front of Bob's laboratory, and then send
it to Bob.  How do the imperfections in Bob's laboratory affect
$p_{*c}$? To quantify this Bob could carry out a complete
tomography of his measurement apparatus, and based on the results
compute what is Alice's best cheating strategy. Here we will make
a simple estimate based on easily accessible parameters.

First of all let us consider the effects of the attenuation $A_T$
during transmission between Alice and Bob's laboratories, of the
attenuation $A_B$ in Bob's  apparatus, and of the efficiency
$\eta$ of his detector. We take these parameters into account by
analysing a fictitious system in which Bob's apparatus is replaced
by a lossless apparatus, and all the attenuation is under Alice's
control, {\it i.e.} $\eta^{fict}=100\%$, $A_B^{fict}=1$, and
$A_T^{fict}=A_T A_B \eta$. This replacement can only help a
cheating Alice. In the fictitious system the state sent by an
honest Alice is $|\pm \alpha_B^{fict}\rangle= |\pm \alpha
\sqrt{A_T A_B \eta}\rangle$.

Second we analyse the effect of finite visibility on the
performance of the   fictitious system just described. Because of
the finite visibility, Bob will not be making a projection onto
the state $|\pm \alpha_B^{fict}\rangle$, but onto slightly
different states. We make the assumption that Bob's apparatus acts
as a passive linear optical system. This implies that the true
states onto which Bob projects are slightly modified coherent
states $|\pm \alpha_B^{fict} + \delta_\pm\rangle$. The deviations
due to $\delta_\pm$ give rise to the optical contribution to the
QBER:
\begin{equation}
QBER_{opt}=\left( |\delta_+|^2 + |\delta_-|^2\right)/2 =q
|\alpha_B^{fict}|^2\ ,
\end{equation}
where $q$, the QBER per photon, can be related to the visibility V
of interferences by $q\simeq (1-V)/2$.
%%%***%%%
(Note that in addition to $QBER_{opt}$, there is another
contribution to the QBER due to the dark counts of the detectors.
The total QBER is the sum of these two contributions:
$QBER=QBER_{opt}+QBER_{dk}$.)

The distance between the two states onto which Bob projects is
given by
\begin{eqnarray}
&&|(+\alpha_B^{fict} + \delta_+)-(-\alpha_B^{fict} + \delta_-)|^2\nonumber\\
&\geq& 4 |\alpha_B^{fict}|^2 - 4 |\alpha_B^{fict}| |\delta_+ -
\delta_-|
%\nonumber\\ &=&
= 4 |\alpha_B^{fict}|^2 (1 - 2 \sqrt{q}) \ .
\nonumber\\
\end{eqnarray}
Inserting this into equation (\ref{p*c}) gives
\begin{equation}
p_{*c}  \leq \frac{1}{2} + \frac{1}{2}\exp\left[ -A_B A_T
\eta\left(1-2 \sqrt{q} \right) \alpha^2\right]\ . \label{p*cB}
\end{equation}
Thus the effect of the imperfections is to replace $\alpha^2$ by
and effective attenuated intensity $A_{B} A_T \eta
\left(1-2\sqrt{q} \right) \alpha^2$.

\subsection{Experimental results}

\begin{figure*}
\includegraphics[width=1\columnwidth]{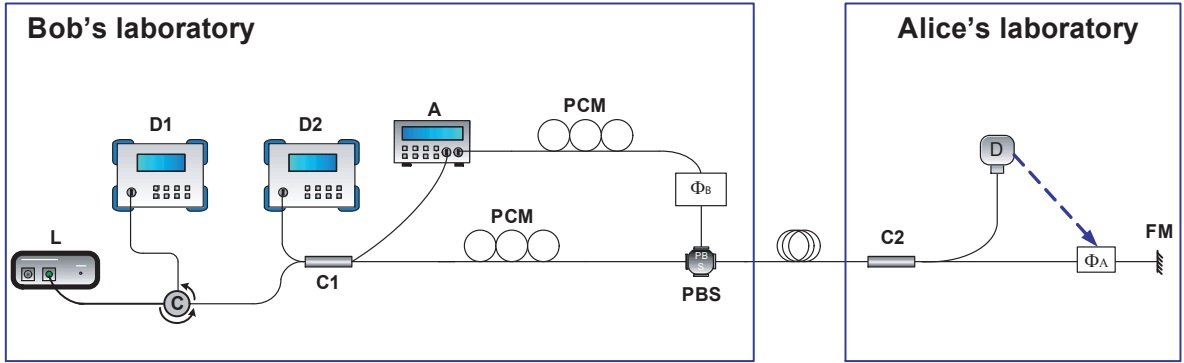}
\caption{Experimental setup: L, picosecond laser; D1, main photon
counter; D2, auxiliary photon counter; C1, 50/50 coupler; A,
attenuator; $\Phi_B$ Bob's phase modulator; PCM, polarization
controller; PBS, polarizing beam splitter; C2, 80/20 coupler; PD,
photodiode; $\Phi_A$, Alice's phase modulator; FM, Faraday
mirror.} \label{fig:schema}
\end{figure*}

Our experimental setup, depicted in Fig. 1, based on the plug and
play system developed for long distance quantum key distribution
\cite{RGGGZ}, is very similar to the one described in \cite{Expt}.
It consists of an all-fiber (standard SMF-28) passively balanced
interferometer, and is therefore well suited to long distance
quantum communication. The protocol begins with Bob producing a
short (300 ps) intense laser pulse at $\lambda = 1.55 \mu m$
(id300 from idQuantique).
%%%***%%%
%containing approximately $3\ 10^6$ photons.
The pulse is split in two by the coupler C1 with equal  reflection
and transmission coefficients $50\%$. The two pulses are delayed
one with respect to the other by 134ns. The pulses are then
recombined on a polarizing beam splitter (PBS) and sent to Alice.
The pulse that propagated along the long arm of the interferometer
is strongly attenuated and will play the role of signal. The pulse
that propagated along the short arm will play the role of local
oscillator (LO). Upon receiving the pulses, Alice splits off part
of them using the coupler C2 and sends this to a photodiode that
triggers her electronics. At Alice's site the pulses are further
attenuated by the different optical elements. They are reflected
by the Faraday mirror. And Alice randomly chooses which phase
$\Phi_A=0,\pi$ to put on the signal pulse using her phase
modulator. The signal Alice sends back to Bob is thus the coherent
state $|\pm \alpha\rangle$ with average photon number
$|\alpha|^2=0.27$.

When the pulses come back to Bob's site, they are sent along the
short  and long arm of the interferometer by the PBS and interfere
at coupler C1. In front of the PBS is a delay line belonging to
Bob which ensures that after the pulses enters Bob's laboratory he
bluehas the time to send to Alice the value of the bit $b$ and
then receive from her the value of $a$. In our experiment the
fiber pigtails of the PBS are sufficient to realize the delay.
Upon receiving the value of $a$, Bob puts the corresponding phase
$\Phi_B=a \pi$ on the LO. This ensures that there should be
destructive interference at the output port that goes to the
circulator and then to detector D1 (id200 from idQuantique). If
detector D1 registers a click, Bob aborts. If it does not click,
the outcome of the coin toss is $c=a\oplus b$. The other output of
coupler C1 is monitored by detector D2, although this is not
directly used in the experiment.

There are in fact two security loopholes in this experiment. The
first arises because Alice does not know the intensity of the
signal pulse she attenuates before sending it back to Bob. Thus in
principle Bob could send her a more intense state than expected,
which would mean that the scalar product of the states prepared by
Alice would be smaller than expected. The second security loophole
arises because Bob does not know the intensity of the pulse he
uses as LO. Thus in principle Alice could send Bob the vacuum
state, both in the signal and LO, and cheat perfectly. Both
loopholes could be closed by having Alice (Bob) monitor the
intensity of the signal (LO) before she (he) attenuates it. This
was not realised in the present setup because the laser pulses
used were not intense enough, but would be possible using more
intense or longer laser pulses as in \cite{Expt}, or by using an
isolator combined with an amplitude modulator as in
\cite{MZG2006}.

\subsubsection{Both parties are honest}
As mentioned above, we performed the experiment with
$|\alpha|^2=0.27$. In a typical series 10000 coins were tossed,
and we obtained 5066 occurrences of $c=1$, 2 occurrences of
$c=\perp$, the other outcomes being $c=0$ (which is consistent
with the statistical uncertainty which should be of order $\sqrt{
5000}= 70$). However we insist that the protocol can be used to
toss a single coin.

We estimate the merit function as follows. The abort probability
is estimated by tossing a large number ($1.5\ 10^5$) of coins with
Alice and Bob both honest:
\begin{equation}
p_{\perp\perp}\simeq 1.40\pm0.37\; 10^{-4} \ .\label{ExpPperp}
\end{equation}
where the error comes from statistical uncertainty.

The transmission losses are assumed to be negligible, $A_T=1$, as
both parties are separated by a few meters of optical fiber. Bob's
detector D1 has a $\eta =10\%$ quantum efficiency. It is gated
using a $2.5\ ns$ gate leading to a dark count probability of
$4.7\ 10^{-5}$. The attenuation of the signal in the optical
elements of Bob's laboratory has been measured to be $A_B\simeq
-6$ dB (which includes the $3$ dB losses at coupler C1 where the
signal and the LO interfere). Visibilities, as measured using an
intense signal, were at least 99.0\% (corresponding to $q=5\
10^{-3}$). By inserting these parameters in equations (\ref{pc*2})
and (\ref{p*cB}) we obtain upper bounds for $p_{*c}$ and $p_{c*}$:
\begin{equation}
p_{*c} \leq 0.9971 \label{ExpP*c} \quad \mbox{and}\quad p_{c*}\leq
0.906
%\label{ExpPc*}
\end{equation}
leading to the lower bound for the merit function:
\begin{equation}
{\cal M} \geq 1.33\ 10^{-4}\ .\label{ExpM}
\end{equation}

This bound  may seem very small. Its value is roughly explained by
noting that the maximal value  in the absence of imperfections is
${\cal M}_{max}=0.021$. The main source of imperfections are the
efficiency of the detectors (10 dB) and the losses in Bob's
apparatus (6 dB). Thus we should reduce the attainable value of
${\cal M}$ by a factor $40$, yielding approximately  equation
(\ref{ExpM}). This argument shows that the simplest way to improve
the experiment would be to use a more efficient detector. It also
shows that the value of ${\cal M}$ is rather robust against small
variations of the experimental parameters. We have computed that
we could keep ${\cal M}$ positive while increasing losses between
Alice and Bob to $A_T\simeq 4.4$ dB (more than 20 km of SMF-28
fiber), all other parameters being kept constant.

\subsubsection{Bob is dishonest}
In order to cheat Bob must estimate the state $|\psi_{a} \rangle$
prepared by Alice so as to correctly guess the value of $a$ before
sending the value of $b$. We implemented a simple  cheating
strategy in which Bob always applies $\Phi_B=0$ on the LO. If
detector D1 clicks Bob assumes that Alice chose $a=1$, whereas if
D1 does not click he assumes $a=0$. Implementing this strategy
yielded the value $p_{1*}=0.505$.
%, which is very close from the statistical deviation
%when both parties are honest.
This very low value is due to the small values of $\eta$ and
$A_B$. Note that a much better cheating strategy, but which was
impossible to implement in our laboratory, would be for Bob to
carry out a homodyne measurement and measure the quadrature that
gives him the best estimate of $a$.

\subsubsection{Alice is dishonest}
As discussed above, when Alice is dishonest her best strategy is
to send a fixed state $|\phi\rangle = N (|+\alpha\rangle +
|-\alpha\rangle)$ to Bob. After receiving $b$ she then sends the
value of $a$ that makes her win the coin toss and hopes that Bob
will not abort. In practice we implemented a strategy where Alice
always sends $|+\alpha\rangle$. Even though this strategy is very
basic, it leads to $p_{*c}=0.9956$, which is very close to the
theoretical maximum equation (\ref{ExpP*c}).

\section{Conclusion}

In conclusion we have studied in detail how the performance of
quantum coin tossing protocols in the presence of imperfections
should be compared to classical protocols. We then reported on a
fiber optics experimental realisation of a quantum coin tossing
protocol. Our analysis shows that in this realisation the maximum
success cheating probabilities for Alice and Bob are respectively
0.9971 and 0.906 when experimental imperfections are taken into
account, which is still better than achievable by any classical
protocol. We implemented this protocol using an all-optical fiber
scheme and tossed a coin whose randomness is higher than
achievable by any classical protocol. Finally we implemented
simple realisable cheating strategies for both Alice and Bob.

After the present work was completed, we learned of a recent
proposal specially designed for carrying out quantum coin tossing
in the presence of losses\cite{BBBG}. Obviously taking into
account losses, in particular those that occur in Bob's apparatus,
was an important consideration when choosing and analysing the
protocol reported here. The protocol reported in \cite{BBBG} seems
more tolerant to loss then ours. Once the effect of other
imperfections (such as finite visibility of interference fringes)
are taken into account, it could be compared to ours using the
merit function ${\cal M}$ introduced above.

\ack
We acknowledge the support of the Fonds pour la formation
\`{a} la Recherche dans l'Industrie et dans l'Agriculture (FRIA,
Belgium); of the Interuniversity Attraction Poles Programme -
Belgian State - Belgian Science Policy under grant IAP6-10; and of
the EU project QAP contract 015848.

\appendix

\section*{Appendix}
\setcounter{section}{1}

Here we provide bounds on the performance of classical coin
tossing protocols  when there is some probability that the
protocol aborts when both parties are honest. We also show that
there exist classical protocols that attain these bounds. We use
the notation and terminology introduced in the main text. The idea
of the following result is to analyse the performance of a
classical protocol with 3 outcomes ({\em i.e.} a classical
protocol in which the parties try to toss a trit.).

{\em Lemma 1: For any correct {\em classical} coin tossing
protocol with three outcomes $0,1,\perp$ we have:
\begin{eqnarray}
(1-p_{0*})(1-p_{*1})&\leq& p_{\perp\perp} \label{rr3}\ ,\\
(1-p_{1*})(1-p_{*0})&\leq& p_{\perp\perp} \label{rr4}\ .
\end{eqnarray}
%where $p_{\perp\perp}$ is the probability that the protocol aborts when both
%parties are honest.
}

{\em Proof of Lemma 1}. We need to introduce some notation.

The protocol consists of $K$ rounds of communication, labeled
$j=1,\ldots,K$.

Denote by $u_j$ the possible states of the protocol at round $j$.

Denote by $w(u_j)$ the probability of reaching state $u_j$ at
round $j$ in an honest execution of the protocol.

Denote by $w(u_{j+1}|u_j)$ the probability that in an honest
execution, the protocol will be in state $u_{j+1}$ at round $j+1$
if it is in state $u_j$ at round $j$.

Denote by $p_{*y}(u_j)$ the maximum probability that if Alice is
dishonest and Bob is honest, then Alice can force Bob to output
$y$ at the end of the protocol if the state at round $j$ is $u_j$.

Denote by $p_{x*}(u_j)$ the maximum probability that if Bob is
dishonest and Alice is honest, then Bob can force Alice to output
$x$ at the end of the protocol if the state at round $j$ is $u_j$.

Introduce the quantity $T_j$ defined by
$$
T_j(x,y)=\sum_{u_j}w(u_j)(1-p_{x*}(u_j))(1-p_{*y}(u_j))
$$

Note that if we take $x=0$ and $y=1$ the initial value ($j=1$) of
$T$ is $T_1(0,1)=(1-p_{0*})(1-p_{*1})$, ie. the left hand side of
eq. (\ref{rr3}).

Note also that at round $K$, when the protocol has ended, $T_K$ is
equal to the sum over the final states of the protocol in an
honest execution of the product of the probabilities that the
output of Alice is not $x$ and that the output of Bob is not $y$.
Thus if we take $x=0$ and $y=1$, then $T_K(0,1)=p_{\perp\perp}$,
ie. the right hand side of eq. (\ref{rr3}).

To complete the proof we  show that $T$ is an increasing function
of $j$, ie. $T_{j+1}\geq T_j$. To this end suppose that at round
$j$ Bob will send some communication to Alice.

Then Alice cannot influence what will happen at round $j$, hence
we have: $p_{*y}(u_j)=\sum_{u_{j+1}}w(u_{j+1}|u_j)
p_{*y}(u_{j+1})$.

Furthermore we have the trivial identity $w(u_{j+1}) = \sum_{u_j}
w(u_{j+1}|u_j) w(u_j)$.

Finally we note that since it is Bob's turn to talk at round $j$,
we have $1-p_{x*}(u_j)\leq 1-p_{x*}(u_{j+1})$ where $u_{j+1}$ is
any state at round $j+1$ that can be obtained from state $u_j$ at
round $j$ in an honest execution.

Inserting these identities into the definition of $T_j$, we
obtained the desired inequality $T_{j+1}\geq T_{j}$.

The proof of eq. (\ref{rr4}) is similar.

{\em End of proof of Lemma 1.}

We have also obtained a partial converse of Lemma 1:
\\
{\em Lemma 2: There exists a correct classical protocol such that
inequality (\ref{rr3}) is saturated, and there exists a correct
classical protocol such that inequality (\ref{rr4}) is saturated.
There also exists a correct classical protocol for which
\begin{eqnarray}
(1-p_{0*})(1-p_{*1})= (1-p_{1*})(1-p_{*0})=\frac{ p_{\perp\perp}
}{2} \ . \label{sat}
\end{eqnarray}}

{\em Proof of Lemma 2.} Let us consider the following protocol:

Round 1: Alice excludes one of the outcomes. That is she chooses
that the outcome of the protocol will be either in $\{0,1\}$ (she
has excluded $\perp$), $\{0,\perp\}$ (she has excluded $1$) or
$\{1,\perp\}$ (she has excluded $0$). She tells her choice to Bob.
If she is honest she chooses randomly among these three
possibilities with a priori probabilities $q_{01}$, $q_{0\perp}$,
$q_{1\perp}$.

Round 2: Bob chooses which of the remaining two outcomes is the
result of the protocol. He tells Alice what is his choice. Thus
for instance if Alice told him that the outcome was in $\{0,1\}$,
Bob can choose that the outcome is either $0$ or $1$, but not
$\perp$. If he is honest he chooses randomly among the two
remaining possibilities with probabilities $q_{0|01}$, $q_{1|01}$;
$q_{0|0\perp}$, $q_{\perp|0\perp}$; $q_{1|1\perp}$,
$q_{\perp|1\perp}$.

It is easy to check that, if the parties are honest, the
probabilities  are:
\begin{eqnarray}
p_{00} &=& q_{0|01} q_{01}+ q_{0|0\perp}q_{0\perp}\nonumber\\
p_{11} &=& q_{1|01} q_{01}+ q_{1|1\perp}q_{1\perp}\nonumber\\
p_{\perp\perp} &=& q_{\perp|0\perp}q_{0\perp}+ q_{\perp|1\perp}q_{1\perp}\ ; %\nonumber\\
\end{eqnarray}
and that, if they are dishonest, the probabilities are:
\begin{eqnarray}
p_{*0}&=&\max \{ q_{0|01},q_{0|0\perp} \}\nonumber\\
p_{*1}&=&\max \{ q_{1|01},q_{1|1\perp} \}\nonumber\\
p_{0*}& =& q_{01}+q_{0\perp}\nonumber\\
p_{1*}&=& q_{01}+q_{1\perp}\ .%\nonumber
\end{eqnarray}

If we choose the parameters such that $q_{0\perp}=q_{1\perp}$,
$q_{0|01}=q_{1|01}=1/2$, $q_{0|0\perp}=q_{1|1\perp}\geq 1/2$,
then the protocol is correct and  eq. (\ref{sat}) is verified.

And if we choose $q_{0\perp}=0$ then we have
$p_{\perp\perp}=q_{\perp|1\perp}q_{1\perp}=(1-q_{1|1\perp})(1-q_{01})$
and $p_{0*}=q_{01}$, $p_{*1}=q_{1|1\perp}$ thus saturating eq.
(\ref{rr3}). Note that by adjusting the remaining free parameter
$q_{0|01}$ one can make the protocol correct.

Similarly one can saturate inequality (\ref{rr4}).

{\em End of proof of Lemma 2.}

%%%%%%%%%%%%%%%%%%%%%%%%%%%%%%%%%%%%%%%%%%%%%%%%%%%%%%%%%%%%%%%%%%%%%%%%%%%%%%%%%%%%%%%%%%%

%%%%%%%%%%%%%%%%%%%%%%%%%%%%%%%%%%%%%%%%%%%%%%%%%%%%%%%%%%%%%%%%%%%%%%%%%%%%%%%%%%%%%%%%%%%%%%%%%%%%%

\end{document}